\documentstyle[preprint,revtex]{aps}
\begin{document}
\begin{title}Doping Dependence of the Chemical Potential in
 cuprate high T$_c$ superconductors II:
  (Bi,Pb)$_{2}$Sr$_{2}$Ca$_{2}$Cu$_{3}$%
  O$_{10+\mbox{\small\protect\boldmath $\delta$}}$
\end{title}
\draft
\author{G. Rietveld$^{a,b}$, S. J. Collocot$^{c}$,D. van der Marel$^{a,d}$}
\begin {instit}
  Delft University of Technology, Department of Applied Physics,\hfill
  Lorentzweg 1, 2628 CJ Delft, The Netherlands.$^{a}$\hfill\\
  NMI-Van Swinden Laboratorium, Department of Electrical
  Standards,\hfill
  P.O. Box 654, 2600 AR\ \ Delft, The Netherlands.$^{b}$\hfill\\
  Affiliation of Steve Collocot.$^{c}$\hfill\\
  Materials Science Centre, Solid State Physics Laboratory\hfill
  University of Groningen,\hfill
  Nijenborgh 4, 9747 AG Groningen, The Netherlands$^d$
\end{instit}
\date{\today}
\begin{abstract}
  Using X-ray photoelectron spectroscopy, a systematic study is performed
  of the doping dependence of the chemical potential $\mu$ in two systems
  of (Bi,Pb)$_2$Sr$_2$Ca$_2$Cu$_3$O$_{10+\delta}$.
  The doping is varied by changing the O content
  of the sample. The measured shifts of the chemical potential
  are compared with present models for the doping behaviour of $\mu$
  in high-$T_{\rm c}$ materials.
\end{abstract}

\newpage
\section{Introduction}

This paper is a sequel to Ref. [\cite{gert.a}], where we
studied the doping dependence
  of the chemical potential of La$_{2-x}$Sr$_x$CuO$_4$ as a function of Sr
  content.
  In the present paper we present measurements of the behaviour of the
  chemical potential as a function of doping in
  (Bi,Pb)$_2$Sr$_2$Ca$_2$Cu$_3$O$_{10+\delta}$. Here, the doping is varied
  by annealing a sample in different oxygen atmospheres.
  As in the previous article, we will first discuss preparation
  and characterization of the sample, and subsequently the XPS
  measurements and the behaviour of the chemical potential in this material.

\section{Preparation and characterization}
The family of Bi$_2$Sr$_2$Ca$_n$Cu$_{n+1}$O$_{2n+6+\delta}$ materials is
particularly suited for XPS
  studies because they are relatively inert and not sensitive to
  moisture and other contaminants in the environment.
  On the other hand it is difficult to prepare single phase
  material. The early {Bi-Sr-Ca-Cu-O} samples predominantly
  contained the $n=1$ material, with certain fractions of the
  $n=0$ and $n=2$ phases, depending on the preparation.
  Later it was found that the addition of Pb stabilizes the
  structure. Especially the $n=2$ material, used in this study,
  can not be made single phase without this addition.

Our (Bi,Pb)$_2$Sr$_2$Ca$_2$Cu$_3$O$_{10+\delta}$ sample is prepared
by a two-step ceramic route
  starting from Bi$_2$O$_3$, PbO, SrCO$_3$, CaCO$_3$ and CuO
  powders of approximately 99~\% purity. A Pb-free Bi-Sr-Ca-Cu-O
  precursor was first prepared by calcination at 800 and
  840~$^{\circ}$C. The required amount of PbO was then added, and
  a final sintering was performed in air at 860~$^{\circ}$C, and then
  850~$^{\circ}$C for a total of 80 hours \cite{Collocott90}.
  The nominal composition of the sample, as determined from the
  starting composition of the oxides in the preparation, is
  Bi$_{1.84}$Pb$_{0.34}$Sr$_{1.91}$Ca$_{2.03}$Cu$_{3.06}$O$_{10+\delta}$.

The sample was given four different surface preparation treatments
  in our UHV system, resulting in four different values of the
  oxygen content as determined from the relatife intensities of the
  XPS lines. Each time the sample surface was freshly scraped {\em in situ}
  with a
  sapphire plate, before it was heated and/or exposed to ozone.
  The pressures and
  temperatures used in the ozone anneals are
  typical for in situ growth of high-quality YBa$_2$Cu$_3$O$_{7-\delta}$
  thin films
  in our system \cite{Appelboom92}. The freshly scraped surface always
  exhibited
  almost the same set of XPS spectra, from which we conclude that
  the changes in oxygen content due to heat/ozone treatment are
  the strongest in a top layer thinner than the 'scraping' depth
  (about 0.1 mm).

Treatment A was to transport the sample immediately to the measurement
  chamber under UHV.\\
  In treatment B the sample was heated to 330~$^{\circ}$C in $3 \times
  10^{-4}$~mbar ozone for 15 minutes, and the ozone delivery
  was stopped during cool-down.\\
  Treatment C was to anneale the sample
  at 450~$^{\circ}$C in $6 \times 10^{-4}$~mbar ozone for 20 minutes,
  and cooled down to room temperature in the same ozone atmosphere.\\
  Treatment D consisted of heating to 100~$^{\circ}$C in UHV for 15
  minutes. The lower temperature for this vacuum anneal was chosen as
  the (Bi,Pb)$_2$Sr$_2$Ca$_2$Cu$_3$O$_{10+\delta}$ material is known
  to decompose when heated to high temperatures at low oxygen pressures.

We have characterized the sample with X-ray diffraction and resistance
  measurements it was introduced in the UHV system,
  and after the series of surface treatments and XPS measurements
  were completed.

The X-ray powder diffraction pattern of the as-prepared
  sample is displayed in Fig.~\ref{fig:mux.bi2xrd}.
  All peaks can be assigned to the $n=2$ material.
  There is no indication of the presence of an $n=1$ phase,
  which is the most frequently found
  impurity in Bi$_2$Sr$_2$Ca$_2$Cu$_3$O$_{10+\delta}$ samples.
  This is best visible from the narrow $(00\underline{14})$ peak,
  which is considerably broadened by the $(00\underline{12})$
  peak of Bi$_2$Sr$_2$CaCu$_2$O$_{8+\delta}$ if this material is present.
  A least squares fit to the peak positions using a tetragonal
  unit cell, gives for the lattice parameters $a= 5.403 \pm
  0.002$~\AA\ and $c=37.02 \pm 0.02$~\AA, which are typical
  values found for this material.

The resistance was measured with a standard 4-terminal
  technique. The data shown in Fig.~\ref{fig:mux.bi2rt}
  are plotted relative to the resistance at 300~K.
  For the as-prepared sample the resistance curve is almost linear
  above $T_{\rm c}$. The onset critical temperature is 110~K. Zero
  resistivity is obtained at 99--103~K, slightly depending on the
  magnitude of the measuring current due to the granularity of
  the material. By comparison with the results of
  Ulm{\em et al.}\cite{Um92}, who studied the $c/a$ ratio and
  the transportproperties of (Bi,Pb)$_2$Sr$_2$Ca$_2$Cu$_3$O$_{10+\delta}$
  after anneals at 700~$^{\circ}$C in different
  oxygen pressures, we conclude,
  that the as-prepared sample is close to being optimally doped.

After completing the XPS analysis following
  the last surface treatment (annealing in vacuum),
  the X-ray diffraction pattern (Fig.~\ref{fig:mux.bi2xrd}).
  indicates that the material is still single phase.
  The $a$-axis value is unaltered within our experimental
  accuracy of $0.02\AA$, but the $c$-axis
  is slightly expanded to $c=37.10 \pm 0.02$~\AA.

The resistance curve above $T_{\rm c}$
  has become non-linear, and the onset $T_{\rm c}$ is lowered
  to 105~K. The sample
  is fully superconducting below 89~K. Comparing these results with
  the oxygen doping study of Um{\em et al.}\cite{Um92} we see
  that the composition has moved in the direction of the metal-insulator
  transition. The smaller Cu~$2p_{3/2}$ as determined with XPS
  (discussed in the following section)
  main peak \cite{Veenendaal93} also points in this direction.
  It is important to notice, that the changes in the oxygen
  concentration at the sample surface are substantially larger
  than in the bulk, as follows from the intensities of the
  XPS lines. As the changes in $\mu$ are also measured
  with the same surface sensitive probe, we will use XPS
  as a quantitative measure of both the oxygen stoichiometry as
  well as the doping dependence of the chemical potential.

\section{Photoelectron Spectra}
After each treatmet of the sample surface in a UHV chamber using the
  different precedures described above, the sample was immediately
  transported to the analysis chamber without breaking the ultrahigh vacuum.

The core level and valence band spectra are displayed
  in Fig.~\ref{fig:mux.bi2xps}.
  All spectra were measured with Mg~$K$$\alpha$ radiation and
  an analyser resolution of $0.4$~eV.
  The carbon contamination was checked with $1$~eV resolution,
  which gives a strongly increased sensitivity.
  The binding energies of the Bi~$4f_{7/2}$, Pb~$4f_{7/2}$,
  Sr~$3d_{5/2}$, Ca~$2p_{3/2}$, Cu~$2p_{3/2}$ and O~$1s$ peaks
  of the as-prepared sample are 158.5, 137.4, 132.4, 345.0,
  933.2 and 528.7~eV respectively.

The Bi~$4f$ spectrum consists of two multiplet-split peaks, which
  have a slight asymmetry towards higher binding energy, which can
  be taken as an indication of a coupling of the core levels to
  excited states in the metallic or semiconducting BiO blocking layers.
  Also the Pb~$4f_{7/2}$ spectrum, shown in
  Fig.~\ref{fig:mux.bi2xps} together with the Sr~$3d$
  spectrum, has an asymmetric line shape.
  It is known that Pb substitutes for Bi in the
  Bi$_2$Sr$_2$Ca$_2$Cu$_3$O$_{10+\delta}$
  superconductor. Unlike Golden {\it et al.} who find a larger
  asymmetry for Pb than for Bi \cite{Golden89}, the $4f$
  line shapes in our case are exactly the same.
  Although the spectra can be fitted to a Doniach-\v{S}unji{\'c}--like
  line shape \cite{Doniach70}, due to
  limitations imposed by the life-time broadening of the XPS lines as
  well as instrumental broadening, a small
  semiconductor gap may exist in the electron-hole shake-up tail.
  We observe that the lineshape is insensitive to the oxygen content.
  Hence, if the asymmetry should be interpreted
  as a mixture of two valencies
  (Bi$^{3+}$ and Bi$^{5+}$) of Bi in this compound
  \cite{Hillebrecht89,Fujimori89}, the average
  valency of Bi has to be independent on the degree of oxygenation.
  The shift of the Bi~$4f$ level (also found
  in the Bi~$5d$ spectra [Fig.~\ref{fig:mux.bi2xps}]) is much larger
  than the shift of the Cu~$2p$ spectrum. Also the core levels of Pb
  and Sr (which is next to the BiO planes) show a larger than
  average shift [Fig.~\ref{fig:mux.bi2shift}].
  This indicates that the extra oxygen is
  introduced near the Bi- and Pb-atoms, supporting the point of
  view that the doping of the CuO$_2$ planes is controlled through
  the oxygen stoichiometry of the blocking layers. Together with
  the fact that the Bi line shapes remain unaffected, this indcates that
  the Bi core levels experience a decrease
  in binding energy due to the electrostatic potential surrounding
  the O$^{2-}$ ions. In the Bi$_2$Sr$_2$Ca$_{1-x}$Y$_x$Cu$_2$O$_{8+\delta}$
  the Y${3+}$ ions replace the Ca between the CuO$_2$ planes, resulting
  in a shift of the Bi core level which is {\em smaller}
  than the average,
  as has also been observed experimentally.\cite{Itti91,Veenendaal93}

The Sr~$3d$ spectrum of Sr containing high-$T_{\rm c}$ superconductors
  is discussed extensively in Ref.\cite{gert.a}. The spectrum
  of the (Bi,Pb)$_2$Sr$_2$Ca$_2$Cu$_3$O$_{10+\delta}$ material
  resembles that of
  Bi$_2$Sr$_2$CaCu$_2$O$_{8+\delta}$\cite{gert.a} . Thus either it
  consists of two components, one of them due to Sr-Ca disorder,
  or it has only one chemical component, with an asymmetric line
  shape.

A single doublet is seen in the Ca~$2p$ spectrum, very similar
  to the early spectrum of Steiner {\it et al.} \cite{Steiner88}.
  Later spectra of highly oriented and of single crystalline
  Bi$_2$Sr$_2$CaCu$_2$O$_{8+\delta}$ material showed at least
  two spin-orbit-split pairs
  marking Sr-Ca disorder \cite{Hill88,Hillebrecht89}.
  The unusual background in our spectrum is caused by Cu~$LMM$
  Auger peaks.

In the Cu~$2p_{3/2}$ spectrum, there is the familiar double
  structure typical for Cu$^{2+}$. After treatment A (only scraping)
  the intensity ratio of satellite and main peak is 0.33,
  somewhat lower than for {\it e.g.} La$_{2-x}$Sr$_x$CuO$_4$.
  This ratio is almost
  independent of oxygen treatment. After treatment D (annealing in vacuum)
  the main peak became somewhat narrower.

The O $1s$ core level is relatively broad, but in contrast
  with that of La$_{2-x}$Sr$_x$CuO$_4$ and YBa$_2$Cu$_3$O$_{7-\delta}$,
  almost free of a high binding energy
  shoulder. Similar oxygen core level spectra were measured
  on the Bi$_2$Sr$_2$CaCu$_2$O$_{8+\delta}$ material
  \cite{Fujimori89,Shen88a,Meyer88a}.
  The large width of the peak is probably caused by slightly
  different binding energies of the different types of oxygen in the
  material, {\it e.g.} as predicted in bandstructure calculations
  \cite{Massidda88}. Afer treatment D a weak signal of carbon
  were present. All other treatments resulted in
  carbon-free surfaces.

\section{Doping Dependence of the Chemical Potential}
Clearly visible in Fig.~\ref{fig:mux.bi2xps} is the systematic shift
  of the spectra as a function of oxygen treatment. For a useful
  plot of this shift, a quantitative determination of the oxygen
  content is necessary. Using the
  calculated cross-section for each level \cite{Scofield76} and
  assuming a random distribution of the elements in the sample,
  we find that the surface composition after scraping (treatment A) is
  Bi$_{2.2}$Pb$_{0.26}$Sr$_{2.0}$Ca$_{1.6}$Cu$_{2.8}$O$_{9.4}$,
  which is close to the nominal value in the bulk.
  In this chemical formula we scaled the cation content
  such as to make the total charge on the cations equal to
  $+20$, taking Bi$^{3+}$, Pb$^{2+}$, Sr$^{2+}$,
  Ca$^{2+}$, Cu$^{2+}$ as the relevant formal
  valencies \cite{Hegde88}. We define $\delta$ as the difference between
  the number of oxygen atoms based on the XPS peak intensities,
  and the number of O$^{2-}$ ions required for total charge compensation
  of the cations. This we interprete as a quantitative
  measure of the {\em changes} in oxygen content.
  After treatments D, A, B, and C this
  is $-0.9$, $-0.6$, $+0.2$, and $+1.3$ respectively.
The result of the shift in each core level versus $\delta$ is given
  in Fig.~\ref{fig:mux.bi2shift}.
  As before, the shifts in the Cu and valence band spectrum are
  determined from the leading edge of the spectra---for all other
  peaks, the peak position is used.
  Over the studied range of $\delta$-values, the average shift is
  reasonably linear with a small upturn after annealing in vacuum.
  By varying the source power during the XPS
  measurements we checked that the observed shifts in the spectra are
  not due to charging of the sample.

Similar to La$_{2-x}$Sr$_x$CuO$_4$, the valence band follows the average
  core level shift quite well, indicating that the observed shift is due to a
  variation of the chemical potential. We see from
  Fig.~\ref{fig:mux.bi2shift} that there is a clear shift
  in chemical potential when increasing the oxygen content
  toward the metallic regime.
  The shift in $\mu$ as a function of doping is larger
  than in La$_{2-x}$Sr$_x$CuO$_4$ \cite{gert.a} and in
  the Nd$_{2-x}$Ce$_x$CuO$_4$ compounds \cite{allen}. We observe a
  stronger shifting of $mu$ upon approacing the MI transition, which
  is in general agreement with previous results with $Y$-doping in
  Bi$_2$Sr$_2$Ca$_{1-x}$Y$_x$Cu$_2$O$_{8+\delta}$.\cite{Veenendaal93,Shen91}
  This inconsistency
  would be removed if in the doped La and Nd copper oxides an additional
  band is formed within within the charge-transfer gap, where the Fermi
  level remains pinned upon doping with holes or electrons. In principle
  such a band might exist due to the close proximity of the (disorderded)
  layer of dopants to the metallic CuO$_2$ sheets. The doping layers
  are indeed further away in the BSCSO system, which would then
  explain the absence (or a different binding energy) of such a band in
  these materials. The question whether such a band could
  indeed exist, poses an interesting theoretical challenge involving both a
  quantative theory of the correlation induced insulating gap, and
  also a quantitative description of states within the
  gap induced by the dopant atoms.

\section{Conclusions}
  We studied the doping dependence of the chemical potential in
  (Bi,Pb)$_2$Sr$_2$Ca$_2$Cu$_3$O$_{10+\delta}$ as a function of O
  concentration
  $\delta$ with X-ray photoelectron spectroscopy. In the metallic
  regime the shift is
  considerably larger than for the La$_{2-x}$Sr$_x$CuO$_4$ system.
  We observe that the chemical potential shifts more strongly, when we
  come closer to the metal-insulator transition by means of oxygen depletion.
  Differences between the Bi and Cu core level shifts indicate
  that the oxygen enters the crystal structure closer to the Bi sites than to
  the Cu atoms.

\section{Acknowledgements}
This investigation was supported by the Netherlands Foundation for
Fundamental Research on Matter (FOM), and the Dutch National
Research Program for high-$T_c$ superconductivity (NOP).

  \figure{ X-ray powder diffraction pattern of the
      (Bi,Pb)$_2$Sr$_2$Ca$_2$Cu$_3$O$_{10+\delta}$ sample directly
      after preparation. All peaks can be ascribed to the $n=2$
      material; impurity phases, such as the frequently found
      $n=1$ phase, are absent. The majority of the peaks is
      labeled with their Miller indices. Cu $K$$\alpha$
      radiation with a Ni filter was used in the measurement.
    \label{fig:mux.bi2xrd}}
  \figure{ Temperature dependent resistance measurements of
      (Bi,Pb)$_2$Sr$_2$Ca$_2$Cu$_3$O$_{10+\delta}$
      relative to the resistance at room temperature
      for the as-prepared and the vacuum annealed sample (solid
      and dashed line respectively).
      After vacuum anneal, both the lower $T_{\rm c}$ and the non-linear
      behaviour of the resistance above $T_{\rm c}$ indicate a lower
      quality of the sample.
    \label{fig:mux.bi2rt}}
  \figure{ Core level and valence band spectra of the
      (Bi,Pb)$_2$Sr$_2$Ca$_2$Cu$_3$O$_{10+\delta}$ sample
      as a function of O content.
      From top to bottom: spectra after treatment D, A, B and C.
      Spectra were obtained using Mg $K$$\alpha$ radiation.
      Note the systematic shift in all spectra, indicating
      a variation in chemical potential.
    \label{fig:mux.bi2xps}}
  \figure{ Observed shifts in core level and valence band position of
      (Bi,Pb)$_2$Sr$_2$Ca$_2$Cu$_3$O$_{10+\delta}$ as a function of oxygen
      content $\delta$. Here,
      $\delta$ is the difference between the measured O content
      and that based on the total cation content of the material.
      From left to right on the horizontal axis are data after
      treatment D, A, B. and C.
      The solid line gives the average shift found in the core
      levels. The dashed line follows the change in valence band
      position.
      The dotted band indicates the scatter in the data (with
      exception of Bi, see discussion in the text).
    \label{fig:mux.bi2shift}}

\end{document}